\documentclass[12pt]{article}
\usepackage{epsfig}
\usepackage{amssymb}
\usepackage{amsmath}
\usepackage{amsfonts}
\usepackage{graphicx}
\usepackage{mathrsfs}
\usepackage[dvips]{color}
\usepackage{multirow}

\newcommand{\R}{\mathbb{R}}
\newcommand{\C}{\mathbb{C}}
\newcommand{\Z}{\mathbb{Z}}

\newcommand{\fc}{\mathfrak{c}}

\newcommand{\fz}{\mathfrak{z}}

\newcommand{\fK}{\mathfrak{K}}

\newcommand{\fS}{\mathfrak{S}}

\newcommand{\bb}{\mathbf{b}}

\newcommand{\bbe}{\mathbf{e}}

\newcommand{\bk}{\mathbf{k}}

\newcommand{\bbr}{\mathbf{r}}
\newcommand{\bx}{\mathbf{x}}

\newcommand{\bcA}{\boldsymbol{\cA}}

\newcommand{\bH}{\mathbf{H}}
\newcommand{\bI}{\mathbf{I}}

\newcommand{\bM}{\mathbf{M}}

\newcommand{\bU}{\mathbf{U}}

\newcommand{\cA}{{\mathcal{A}}}

\newcommand{\cG}{\mathcal{G}}

\newcommand{\cK}{\mathcal{K}}

\newcommand{\cM}{\mathcal{M}}

\newcommand{\be}{\begin{equation}}
\newcommand{\ee}{\end{equation}}
\newcommand{\bea}{\begin{eqnarray}}
\newcommand{\eea}{\end{eqnarray}}
\newcommand{\nn}{\nonumber}

\newcommand{\ed}{\end{document}}

\newcommand{\bi}{\begin{itemize}}
\newcommand{\ei}{\end{itemize}}

\newcommand{\bce}{\begin{center}}
\newcommand{\ece}{\end{center}}

\newcommand{\sT}{\mathscr{T}}

\newcommand{\RE}{{\rm Re}}
\newcommand{\IM}{{\rm Im}}

\newcommand{\bcK}{{\boldsymbol{\cK}}}

\begin{document}

\title{Exact Solution of the Two-Dimensional Scattering Problem for a Class of $\delta$-Function Potentials Supported\\ on Subsets of a Line}

\author{Farhang Loran\thanks{E-mail address:loran@cc.iut.ac.ir}
~and Ali~Mostafazadeh\thanks{E-mail address:
amostafazadeh@ku.edu.tr}\\[6pt]
$^*$Department of Physics, Isfahan University of Technology, \\ Isfahan 84156-83111, Iran\\[6pt]
$^\dagger$Departments of Mathematics and Physics, Ko\c{c} University,\\  34450 Sar{\i}yer,
Istanbul, Turkey}

\date{ }
\maketitle

\begin{abstract}

We use the transfer matrix formulation of scattering theory in two-dimensions to treat the scattering problem for a potential of the form $v(x,y)=\zeta\,\delta(ax+by)g(bx-ay)$ where $\zeta,a$, and $b$ are constants, $\delta(x)$ is the Dirac $\delta$ function, and $g$ is a real- or complex-valued function. We
map this problem to that of $v(x,y)=\zeta\,\delta(x)g(y)$ and give its exact and analytic solution for the following choices of $g(y)$: i) A linear combination of $\delta$-functions, in which case $v(x,y)$ is a finite linear array of two-dimensional $\delta$-functions; ii) A linear combination of $e^{i\alpha_n y}$ with $\alpha_n$ real; iii) A general periodic function that has the form of a complex Fourier series. In particular we solve the scattering problem for a potential consisting of an infinite linear periodic array of two-dimensional $\delta$-functions. We also prove a general theorem that gives a sufficient condition for different choices of $g(y)$ to produce the same scattering amplitude within specific ranges of values of the wavelength $\lambda$. For example, we show that for arbitrary real and complex parameters, $a$ and $\fz$, the potentials $\fz\sum_{n=-\infty}^\infty\delta(x)\delta(y-an)$ and $a^{-1}\fz\delta(x)[1+2\cos(2\pi y/a)]$ have the same scattering amplitude for $a< \lambda\leq 2a$.
\vspace{2mm}



\end{abstract}

\section{Introduction}

Recently we have developed a transfer-matrix formulation of scattering theory in two and three dimensions that allows for an exact solution of the scattering problem for the $\delta$-function potential in two and three dimensions, \cite{pra-2016}. This is particularly remarkable, because unlike the standard approach based on the Lippmann-Schwinger equation, it does not lead to a singular result that would require a renormalization scheme \cite{henderson-1997,henderson-1998,Camblong-2001a,camblong-2002}.  In the present paper, we study the utility of this approach in solving the scattering problem for the potentials of the form	
	\be	
	v(x,y)=\zeta\,\delta(ax+by)g(bx-ay).
	\label{v=1}
	\ee
Here $\zeta,a$, and $b$ are real parameters and $g:\R\to\C$ is a piecewise continuous function having a Fourier transform. The potential (\ref{v=1}) describes a singular interaction that is localized on a subset $S$ of the line $ax+by=0$ in $\R^2$, namely
    \[ S:=\left\{
    \begin{array}{ccc}
    \left\{(0,y)\in\R^2~|~g(-ay)\neq 0\;\right\} & {\rm for} & b=0,\\[6pt]
    \left\{(x,-\frac{ax}{b})\in\R^2~|~g((\frac{a^2}{b}+b)x)\neq 0\;\right\} & {\rm for} & b\neq 0.
    \end{array}\right.\]

Suppose that an incident plane wave with wave vector
    \be
    \bk_0=k(\cos\theta_0\bbe_x+\sin\theta_0\bbe_y)
    \label{incident-wv}
    \ee
scatters off a potential of the form~(\ref{v=1}), where $k$ is the wavenumber, $\theta_0$ is the incidence angle, and $\bbe_j$ is the unit vector along the $j$-axis. If $b\neq 0$, we can rotate the coordinates by an angle $\varphi:={\rm arctan}(-a/b)$ that transforms $x$ and $y$ to $x':=\sin\varphi\,x-\cos\varphi\,y$ and $y':=\cos\varphi\,x+\sin\varphi\,y$. Using these relations in (\ref{v=1}), we have
	\be
	v(x,y)=v'(x',y'):=\zeta'\delta(x')g'(y'),
	\label{v=2}
	\ee
where $\zeta':=\zeta/\sqrt{a^2+b^2}$ and $g'(y'):=g(\sqrt{a^2+b^2}\: y')$. Equation~(\ref{v=2}) together with rotational symmetry of the Laplacian appearing in the Schr\"odinger equation,
	\be
    [-\nabla^2+v(x,y)]\psi(x,y)=k^2\psi(x,y),
    \label{sch-eq}
    \ee
allow us to identify the scattering of the incident plane wave with incident angle $\theta_0$ by the potential (\ref{v=1}) with the scattering of an incident plane wave with incident angle $\theta_0+|\varphi|-\pi{\rm sgn}(\varphi)/2$ by the potential
	\be
	v(x,y)=\zeta\,\delta(x)g(y).
	\label{v=3}
	\ee
In other words, without loss of generality, we can confine our attention to potentials of the form (\ref{v=3}) which are supported on the following subset of the $y$-axis:
    \[S:=\left\{(0,y)\in\R^2~|~g(y)\neq 0\;\right\}.\]

Two interesting special choice for $g(y)$ are
    \bea
    g(y)&=&\sum_{n=1}^{N}\fc_n \delta(y-a_n),
    \label{multi-delta-g}\\
    g(y)&=&\sum_{n=-N}^{N}\fc_n e^{i\alpha_n y},
    \label{multi-exp-f}
    \eea
where $N$ is a positive integer, $\fc_n$ are real or complex constants, and $a_n$ and
$\alpha_n$ are real parameters. These respectively correspond to the potentials:
    \bea
     v(x,y)&=&\delta(x)\sum_{n=1}^{N}\fz_n \delta(y-a_n),
    \label{multi-delta-v-0}\\
    v(x,y)&=&\delta(x)\sum_{n=-N}^{N}\fz_n e^{i\alpha_n y},
    \label{multi-exp-v-0}
    \eea
where $\fz_n:=\zeta\fc_n$. If there is some positive real parameter $\alpha$ such that $\alpha_n$ is an integer multiple of $\alpha$, (\ref{multi-exp-v-0}) is a periodic function of $y$. Our aim is to obtain an exact solution of the scattering problem for the potentials of the form (\ref{multi-delta-v-0}) and (\ref{multi-exp-v-0}) for any choice of $N$, $\fz_n$, $a_n$, and $\alpha_n$, as well as the class of arbitrary $y$-periodic potentials, which we can express as
	\be
    	v(x,y)=\delta(x)\sum_{n=-\infty}^{\infty}\fz_n e^{in\alpha y}.
   	\label{multi-exp-v-periodc}
   	 \ee
In particular, we offer an exact solution of the scattering problem for the periodic $\delta$-function potentials of the form:
	\[v(x,y)=\fz\,\delta(x)\sum_{n=-\infty}^{\infty} \delta(y-na),\]
where $\fz$ and $a$ are respectively nonzero complex and real parameters. This potential models an infinite periodic linear array of point scatterers in two dimensions that has been considered using standard Green's function methods in the literature. See for example \cite{linton-martin} and references therein.

\section{Transfer-matrix in two dimensions}

Consider a scattering potential $v(x,y)$ with sufficiently fast decay rate such that for $x\to\pm\infty$ the solutions of the Schr\"odinger equation~(\ref{sch-eq}) tend to
    \be
    \frac{1}{2\pi}
	\int_{-k}^k dp\, e^{ipy}\left[A_\pm(p)e^{i\omega(p)x}+
	B_\pm(p) e^{-i\omega(p)x}\right],
    \label{asym}
	\ee
where
    \be
    \omega(p):=\sqrt{k^2-p^2}.
    \label{omega}
    \ee
and $A_\pm(p)$ and $B_\pm(p)$ are coefficient functions vanishing for $p\notin[-k,k]$, i.e., they belong to
	\[\fS:=\{\phi:\R\to\C~|~\phi(p)=0~{\rm for}~|p|>k~\}.\]
By definition, the transfer matrix of $v(x,y)$ is the $2\times 2$ matrix operator $\bM(p)$ satisfying
    \be
    \left[\begin{array}{c}
	A_+(p) \\ B_+(p) \end{array}\right]=\bM(p)\left[\begin{array}{c}
	A_-(p) \\ B_-(p) \end{array}\right].
	\label{M-def}
	\ee
Note that in general its entries, which we denote by $M_{ij}(p)$, are linear operators acting in $\fS$.

This notion of transfer matrix has two remarkable properties \cite{pra-2016}:
    \begin{itemize}
    \item[] Property~1: It contains complete information about the scattering properties of the potential.
    \item[] Property~2:  It shares the composition property of its well-known one-dimensional analog \cite{sanchez}.
    \end{itemize}

To describe Property~1 in more detail, we take the $x$-axis as the scattering axis and recall that for a left-incident wave with wavevector (\ref{incident-wv}), the scattering solutions of (\ref{sch-eq}) have the asymptotic form \cite{adhikari}:
    \be
    \psi(\bbr)=e^{i\bk_0\cdot\bbr}+\sqrt{\frac{i}{kr}}\,e^{ikr} f(\theta)~~~~{\rm as}~r\to\infty,
    \label{psi-left}
    \ee
where $\bbr$ is the position vector with Cartesian coordinates $(x,y)$ and polar coordinates $(r,\theta)$, and $f(\theta)$ is the scattering amplitude. It turns out that the latter is uniquely determined by the transfer matrix $\bM(p)$. To see this, we introduce
    \begin{align}
    &T_-(p):=B_-(p), &&T_+(p):=A_+(p)-A_-(p),
    \label{T-def-L}
    \end{align}
and note that for our left-incident wave,
    \begin{align}
	&A_-(p)=2\pi\delta(p-p_0), && B_+(p)=0,
	\label{scat-L}
	\end{align}
where $p_0$ is the $y$-component of $\bk_0$, i.e.,
    \[ p_0:=\bbe_y\cdot\bk_0=k\sin\theta_0,~~~~~
    \theta_0\in[\mbox{$-\frac{\pi}{2},\frac{\pi}{2}$}].\]
In Ref.~\cite{pra-2016} we derive an explicit expression for the scattering amplitude in terms of $M_{ij}(p)$ for a normally incident wave, i.e., $\theta_0=0$. The application of this approach to an incident wave with an arbitrary incident angle yields
	\begin{align}
	&T_-(p)=-2\pi M_{22}(p)^{-1}M_{21}(p)\delta(p-p_0),
    \label{Tm-L}\\[3pt]
    &T_+(p)=M_{12}(p)T_-(p)+2\pi\big[M_{11}(p)-1\big]\delta(p-p_0),
    \label{Tp-L}\\
	&f(\theta)=-\frac{i k|\cos\theta|}{\sqrt{2\pi}}\times\left\{
    \begin{array}{ccc}
    T_+(k\sin\theta) & {\rm for} & -\frac{\pi}{2}<\theta<\frac{\pi}{2},\\
    T_-(k\sin\theta) & {\rm for} & \frac{\pi}{2}<\theta<\frac{3\pi}{2}.\end{array}\right.
 	\label{f-L}
    \end{align}
This completes our discussion of Property~1.

Property~2 follows from the intriguing observation that, similarly to its one-dimensional analog \cite{ap-2014}, the transfer matrix $\bM(p)$ can be expressed in terms of the evolution operator for an effective non-Hermitian Hamiltonian operator. More specifically let $\bU(x,p)$ be the solution of
    \be
    i\partial_x \bU(x,p)= \bH(x,p) \bU(x,p),~~~~~\bU(-\infty,p)=\bI,
    \ee
where $\bH(x,p)$ is the effective non-Hermitian Hamiltonian operator:
    \begin{align}
    &\bH(x,p):=\frac{1}{2\omega(p)} e^{-i\omega(p)x\boldsymbol{\sigma}_3}
	v(x,i\partial_p)\,\boldsymbol{\cK}\,e^{i\omega(p)x\boldsymbol{\sigma}_3},
    \label{H=}\\
    &\boldsymbol{\cK}:=\boldsymbol{\sigma}_3+i\boldsymbol{\sigma}_2=
    \left[\begin{array}{cc}
    1 & 1\\
    -1 & -1\end{array}\right],
    \end{align}
$\boldsymbol{\sigma}_j$ are the Pauli matrices, $v(x,i\partial_p)$ is the integral operator acting in $\fS$ according to
    \be
    v(x,i\partial_p)\phi(p):=\frac{1}{2\pi}\int_{-k}^k dq\,\tilde v(x,p-q) \phi(q),
    \label{int-op}
    \ee
$\tilde v(x,\fK_y):=\int_{-\infty}^\infty dy\,e^{-i\fK_y y}v(x,y)$ is the Fourier transform of $v(x,y)$ with respect to $y$, and $\bI$ is the $2\times 2$ identity matrix. Then $\bM(p)=\bU(\infty,p)$. In other words, we can express the transfer matrix as the time-ordered exponential:
    \bea
    \bM(p)&:=&\sT\exp\left[-i\int_{-\infty}^\infty dx \bH(x,p)\right],
    \nn\\
    &=&\bI-i\int_{-\infty}^\infty dx_1 \bH(x_1,p)+ (-i)^2\int_{-\infty}^\infty dx_2 \int_{-\infty}^{x_2}dx_1
    \bH(x_2,p)\bH(x_1,p)+\cdots
    \label{M=}
    \eea
where $x$ plays the role of time \cite{pra-2016}.

\section{Solution of the scattering problem for $v(x,y)=\zeta\,\delta(x)g(y)$}

For a potential of the form (\ref{v=3}), the effective Hamiltonian (\ref{H=}) takes the following simple form
    \begin{align}
    &\bH(x,p):=i\zeta\,\delta(x)\cG\,\boldsymbol{\cK},
    \label{H=2}
    \end{align}
where $\cG$ is the integral operator acting in $\fS$ according to
	\be
     \cG\phi (p):=-\frac{i}{2\omega(p)}\:g(i\partial_p)\phi(p)=
    -\frac{i}{4\pi\omega(p)}\int_{-k}^k dq\,\tilde g(p-q) \phi(q).
    \label{F=}
    \ee
Because $\bcK^2$ is the zero matrix, the terms involving products of $\bH(x,p)$ on the right-hand side of (\ref{M=}) vanish, and we find the following exact formula for the transfer matrix.
    \be
    \bM(p)=\bI+\zeta\,\cG\,\boldsymbol{\cK}
    =\left[\begin{array}{cc}
    1+\zeta\cG & \zeta\cG\\
    -\zeta\cG & 1-\zeta\cG\end{array}\right].
    \label{M=2}
    \ee
Because entries of $\bM(p)$ map elements of $\fS$ to elements of $\fS$, according to (\ref{M=2}) the  same should hold for $\cG$, i.e., the domain of $\cG$ consists of integrable elements of $\fS$ such that the right-hand side of (\ref{F=}) belongs to $\fS$.

Using (\ref{M=2}) in (\ref{Tm-L}) and (\ref{Tp-L}), we have
    \bea
    T_+(p)=T_-(p)&=&2\pi\left[(1-\zeta\cG)^{-1}-1\right]\delta(p-p_0),
    \label{T=T}\\
    &=&2\pi\sum_{\ell=1}^\infty \zeta^\ell \cG^\ell\delta(p-p_0).
    \label{T-series}
    \eea
The first equation in (\ref{T=T}) together with (\ref{f-L}) imply
    \be
	f(\theta)=-\frac{i k|\cos\theta|T_+(k\sin\theta)}{\sqrt{2\pi}}
	~~~{\rm for~all}~~~\theta\in\mbox{$[-\frac{\pi}{2},\frac{3\pi}{2})$}.
	\label{Scat-Amp}
	\ee
Eq.~(\ref{T-series}) is clearly suitable for a perturbative calculation of $T_\pm(p)$ and hence the scattering amplitude (\ref{Scat-Amp}).

Next, we recall that according to (\ref{T-def-L}) and (\ref{scat-L}),
	\be
	A_+(p)=T_+(p)+2\pi\delta(p-p_0).
	\label{A-plus}
	\ee
In view of this relation, (\ref{T=T}) is equivalent to the following integral equation for $A_+(p)$
	\be
	(1-\zeta\cG)A_+(p)=2\pi\delta(p-p_0).
	\label{eq-A}
	\ee
It is crucial to note that we seek for the solutions of this equation in $\fS$, i.e., $A_+(p)=0$ and $\cG A_+(p)=0$ for all $|p|>k$.

\section{Exactly Solvable Potentials}

In this section we consider different choices of $g(y)$ for which we can solve Eq.~(\ref{eq-A}) exactly.

\subsection{Finite linear array of $\delta$-function potentials in two dimensions}
\label{Sec4-1}

Consider setting
	\be
	g(y)=\sum_{n=1}^N\fc_n\,\delta(y-a_n),
	\label{f-2d-delta}
	\ee
where $N$ is a positive integer, and $\fc_n$ and $a_n$ are respectively complex and real parameters. Then the potential (\ref{v=3}), which assumes the form
	\be
	v(x,y)=\delta(x)\sum_{n=1}^N\fz_n\,\delta(y-a_n),
	\label{2d-delta}
	\ee
with $\fz_n:=\zeta\,\fc_n$, describes a finite linear array of point interactions in two dimensions. For $N=1$, (\ref{2d-delta}) is the $\delta$-function potential in two dimensions, whose bound state problem for real and negative values of $\fz$ has been studied extensively \cite{henderson-1997,henderson-1998,Camblong-2001a,camblong-2002,Lapidus-1982}. The Dyson series solution of the Lippmann-Schwinger equation for this potential involves divergent terms starting with the second order term. Therefore the standard perturbative solution of this equation encounters serious difficulties.

In order to employ our approach for the computation of the scattering amplitude for potentials of the form (\ref{2d-delta}), we first use (\ref{F=}) and (\ref{f-2d-delta}) to compute $\cG A_+(p)$. Substituting the result in (\ref{eq-A}), we find
    \be
	A_+(p)=2\pi\delta(p-p_0)-\frac{i}{2\omega(p)}\sum_{n=1}^N \fz_n x_n e^{-ia_n p} ,
	\label{eq-071}
	\ee
where $x_n:=\check A_+(a_n)$ and for each $\phi\in\fS$,
    \be
    \check\phi(y):=\frac{1}{2\pi}\int_{-k}^k dp\: e^{ipy}\phi(p)
    \label{inv-F}
    \ee
is the inverse Fourier transform of $\phi(p)$.

Comparing (\ref{A-plus}) and (\ref{eq-071}) we can identify the second term on the right-hand side of (\ref{eq-071}) with $T_+(p)$. Substituting this in (\ref{Scat-Amp}) gives
    \be
    f(\theta)=\frac{-1}{2\sqrt{2\pi}}\sum_{n=1}^N \fz_n  x_n  e^{-ia_n k\sin\theta}.
    \label{delta-f=}
    \ee
This reduces the solution of the scattering problem for the potential (\ref{2d-delta}) to the determination of $x_n$. These generally depend on $k$ and $p_0$ and consequently $\theta_0$.

In order to compute $x_n$, we first calculate the inverse Fourier transform of $A_+(p)$ using (\ref{inv-F}). Because $A_+(p)$ vanishes for $|p|>k$, and it is given by (\ref{eq-071}) for $|p|\leq k$, this calculation gives
    \be
    \check A_+(y)=e^{ip_0y}-\frac{i}{4}\sum_{n=1}^N\fz_n x_n J_0[k(y-a_n)],
    \label{eq-072}
    \ee
where $J_0$ stands for the Bessel-J function of order zero, and we have made use of the identity:
$\int_{-k}^k dp\,e^{iap}/\omega(p)=\pi J_0(ak)$. If we evaluate (\ref{eq-072}) at $y=a_m$ and note that $x_n:=\check A_+(a_n)$, we find the following linear system of equations for $x_n$.
    \be
    \sum_{n=1}^N \cA_{mn}x_n=b_m,
    \label{sys-delta}
    \ee
where $m=1,2,\cdots N$ and
    \begin{align}
    &\cA_{mn}:=\delta_{mn}+\frac{i}{4}\,\fz_n J_0[k(a_m-a_n)],
    && b_m:=e^{ia_m p_0}.
    \label{eq-073}
    \end{align}
Eqs.~(\ref{eq-072}) and (\ref{sys-delta}) are Foldy's fundamental equations of multiple scattering \cite{foldy}. Here they follow from the application of our approach to scattering theory \cite{pra-2016} to potentials of the form (\ref{2d-delta}).
   
The values of $k$ for which the matrix $\bcA$ of coefficients $\cA_{mn}$ of the system (\ref{sys-delta}) is singular, i.e., $\det\bcA=0$, correspond to the spectral singularities \cite{prl-2009} of the potential (\ref{2d-delta}).\footnote{Spectral singularities are energies at which a scattering solution of the Schr\"odinger (or Helmholtz) equation behaves like a zero-width resonance \cite{prl-2009}. For an optical system modeled by a scattering potential, they give the laser threshold condition for the system
\cite{pra-2011a}. For a discussion of the optical realizations of scattering potentials in two dimensions and their spectral singularities, see \cite{pra-2016,prsa-2016} and references therein.} Except for these, $\bcA$ is invertible and we can express the solution of (\ref{sys-delta}) in the form $\bx=\bcA^{-1}\bb$, where $\bx$ and $\bb$ are column vectors with entries $x_m$ and $b_m$, respectively. It is interesting to see that $\bcA$ and therefore spectral singularities do not depend on $p_0$ and hence the incidence angle $\theta_0$.

Let us examine some simple particular cases:
\begin{itemize}
\item[1)] $\delta$-function potential in two dimensions:

For $N=1$, (\ref{2d-delta}) is a delta-function potential localized at the point $(0,a_1)$ of the plane, and because $J_0(0)=1$, (\ref{eq-073}) gives $\bcA=\cA_{11}=1+i\fz_1/4$ and $\bb=b_1=e^{ia_1 p_0}$. Therefore, $\bx=x_1=4e^{ia_1 p_0}/(4+i\fz_1)$. Substituting this in (\ref{delta-f=}) and noting that $p_0=k\sin\theta_0$, we find
	\be
	f(\theta)=-\sqrt{\frac{2}{\pi}}\frac{\fz_1 e^{-ia_1k(\sin\theta-\sin\theta_0)}}{4+i\fz_1}.
	\ee
For $a_1=0$ this expression reproduces Eq.~(25) of Ref.~\cite{pra-2016}.

\item[2)] Double-$\delta$-function potential in two dimensions:

For $N=2$, (\ref{2d-delta}) reads
    \be
    v(x,y)=\delta(x)[\fz_1\delta(y-a_1)+\fz_2(y-a_2)],
    \label{double-delta}
    \ee
and (\ref{eq-073}) gives
    \begin{align}
    &\bcA=\frac{1}{4}\left[\begin{array}{cc}
    4+i\fz_1  & i\fz_2J_0[k(a_1-a_2)]\\
    i\fz_1 J_0[k(a_1-a_2)] & 4+i\fz_2\end{array}\right],
    &&\bb=\left[\begin{array}{c}
    e^{ia_1p_0}\\
    e^{ia_2p_0}\end{array}\right].
    \end{align}
In particular, the spectral singularities of (\ref{double-delta}) are given by the real values of $k$ for which
    \be
    \det\bcA=\frac{1}{16}\left\{J_0[k(a_1-a_2)]^2-1\right\}\fz_1\fz_2+\frac{i}{4}(\fz_1+\fz_2)+1
    \ee
vanishes.\footnote{This cannot happen for complex-conjugate and in particular real coupling constants $\fz_1$ and $\fz_2$, because $J_0$ is an entire function and $|J_0(x)|<1$ for $x>0$.} We can easily invert $\bcA$, determine $\bx$, and use (\ref{delta-f=}) to compute $f(\theta)$. This results in
    \bea
    f(\theta)&=&\frac{-1}{8\sqrt{2\pi}\det\bcA}\Big\{
    \fz_1(4+i\fz_2)e^{-ia_1k(\sin\theta-\sin\theta_0)}+
    \fz_2(4+i\fz_1)e^{-ia_2k(\sin\theta-\sin\theta_0)}\nn\\
    &&-i\fz_1\fz_2J_0[k(a_1-a_2)]\left[
    e^{-ik(a_1\sin\theta-a_2\sin\theta_0)}+
    e^{-ik(a_2\sin\theta-a_1\sin\theta_0)}\right]\Big\}.
    \label{f=077}
    \eea

For a pair of identical $\delta$-functions separated by a distance $a$ and localized at  $(0,\pm a/2)$, we have $\fz_1=\fz_2=:\fz$ and $a_1=-a_2=a/2$. In this case, (\ref{f=077}) reduces to
	\be
	f(\theta)=f_-(\fz,k)\cos\left[\mbox{$\frac{ak}{2}(\sin\theta-\sin\theta_0)$}\right]+
	f_+(\fz,k)\cos\left[\mbox{$\frac{ak}{2}(\sin\theta+\sin\theta_0)$}\right],
	\label{f=078}
	\ee
where
	\begin{align}
	&f_-(\fz,k):=\frac{4\fz(4+i\fz)}{\sqrt{2\pi}\Delta(\fz,k)},
	&&f_+(\fz,k):=\frac{-4i\fz^2J_0(ak)}{\sqrt{2\pi}\Delta(\fz,k)},\nn
	\end{align}
	\be
	\Delta(\fz,k):=[1-J_0(ak)^2]\fz^2-8i\fz-16.\nn
    	\ee
Spectral singularities of the potential (\ref{double-delta}) are determined by $\Delta(\fz,k)=0$, which is equivalent to
	\be
	\fz=\frac{4i[1\pm\sqrt{2-J_0(ak)^2}]}{1-J_0(ak)^2}.
	\ee
Because $J_0(x)<1$ for $x>0$, this equation shows that spectral singularities arise only for imaginary values of $\fz$.

For a normally incidence wave, $\theta_0=0$ and (\ref{f=078}) simplifies to
        \be
	f(\theta)=[f_-(\fz,k)+f_+(\fz,k)]\cos\left(\mbox{$\frac{ak\sin\theta}{2}$}\right).
	\label{f=079}
	\ee
In particular, the intensity of the scattered wave, which is proportional to $|f(\theta)|^2$, vanishes for
$\sin\theta=\pm\pi/ak=\pm\lambda/2a$, where $\lambda:=2\pi/k$ is the wavelength of the incident wave. In particular $|f(\theta)|^2>0$ unless of $\lambda<2a$. In this case, the intensity of the scattered wave vanishes along the four directions given by $\theta=\pm{\rm arcsin}(\frac{\lambda}{2a}),\pi\mp{\rm arcsin}(\frac{\lambda}{2a})$. At a spectral singularity the potential emits purely outgoing waves along all directions except these.

\end{itemize}

\subsection{Potentials of the form $\delta(x) \sum_{n=-N}^N\fz_n e^{i\alpha_n y}$}
\label{Sec4-2}

Consider the potentials:
    \be
    v(x,y)=\delta(x)\sum_{n=-N}^{N}\fz_n e^{i\alpha_n y},
    \label{multi-exp-v}
    \ee
which correspond to the choice (\ref{multi-exp-f}) for $g(y)$. For this choice (\ref{F=}) gives
    \be
    \cG A_+(p)=\frac{-i}{2\omega(p)}\sum_{n=-N}^N\fc_n A_+(p-\alpha_n).
    \label{F=3}
    \ee
Since $N$ is arbitrary and the coefficients $\fc_n$ can be zero, without loss of generality we can set $\alpha_0=0$ and demand that $\alpha_{-n}=-\alpha_n<0$ whenever $n\neq 0$.
By inspecting the consequences of employing (\ref{F=3}) in (\ref{eq-A}) and noting that $A_+\in\fS$, we have arrived at the following ansatz for the solution of (\ref{eq-A}):
    \be
    A_+(p)=\sum_{\vec m\in\cM}x_{\vec m}\:\delta(p-p_0-\vec m\cdot\vec\alpha),
    \label{ansatz}
    \ee
Here $x_{\vec m}$ are undetermined complex coefficients, $\vec\alpha:=(\alpha_1,\alpha_2,\cdots,\alpha_N)\in\R^N$, $\alpha_n$ are the parameters of the potential (\ref{multi-exp-v}), $\cM$ is the set of $N$-tuples of integers $\vec m:=(m_1,m_2,\cdots,m_N)\in\Z^N$ such that
	\be
 	|\vec m\cdot\vec\alpha+p_0|\le k,
	\label{bound}
 	\ee
and $\vec m\cdot\vec\alpha:=\sum_{n=1}^Nm_n\alpha_n$.
Notice that $k$ and $p_0$ are input parameters of the scattering problem. For each value of these parameters, (\ref{bound}) puts an upper bound on the possible choices for $\vec m$. This shows that $\cM$ is a finite subset of $\Z^N$. We arrive at the same conclusion by noting that because $|p_0|\leq k$, (\ref{bound}) implies
    \be
    |\vec m\cdot\vec\alpha|\leq 2k.
    \label{condi}
    \ee

Inserting (\ref{ansatz}) in (\ref{eq-A}) and using (\ref{F=3}) we are led to an equation that we can put in the form
	\be
	\sum_{\vec \ell\in\cM}\Big(\sum_{\vec m\in\cM}\cA_{\vec\ell,\vec m}\,x_{\vec m}
	-b_{\vec \ell}\Big)\delta(p-p_0-\vec\ell\cdot\vec\alpha)=0.
	\label{eq11}
	\ee
Here for all $\vec\ell,\vec m\in\cM$,
	\begin{align}
	&\cA_{\vec\ell,\vec m}:=\delta_{\vec\ell,\vec m}+
	\frac{i}{2\omega(p_0+\vec\ell\cdot\vec\alpha)}
	\sum_{n=-N}^N\fz_n\,\delta_{\vec\ell,\vec m+\vec e_n},
	&&b_{\vec\ell}:=2\pi\delta_{\vec\ell,\vec e_0}\;,
	\label{L-b}
	\end{align}
	\begin{align}
	&\delta_{\vec\ell,\vec m}:=\left\{\begin{array}{ccc}
	1 & {\rm for} & \vec\ell=\vec m,\\
	0 & {\rm for} & \vec\ell\neq\vec m,\end{array}\right.
	&& \vec e_n:=\left\{\begin{array}{ccc}
	(0,0,0,\cdots,0) & {\rm for} & n=0,\\
	(\underbrace{0,0,\cdots, 0}_{|n|-1},{\rm sgn}(n),0,\cdots,0) & {\rm for} & n\neq 0.
	\end{array}\right.
	\end{align}
Because $\delta(p-p_0-\vec m\cdot\vec\alpha)$ are linearly independent, (\ref{eq11}) is equivalent to the following linear system of equations for $x_{\vec m}$:
	\be
	\sum_{\vec m\in\cM}\cA_{\vec\ell,\vec m}~x_{\vec m}=b_{\vec\ell}.
	\label{sys}
	\ee
Solving this system we can determine $A_+(p)$ and use (\ref{A-plus}) and (\ref{Scat-Amp}) to obtain $T_+(p)$ and $f(\theta)$, respectively. In fact, it is not difficult to show that
    \be
    f(\theta)=\frac{-i}{\sqrt{2\pi}}
    \sum_{\vec m\in\cM}
     y_{\vec m}
    [\delta(\theta-\theta_{\vec m})+\delta(\theta+\theta_{\vec m}-\pi)] ,
    \label{exp-f}
    \ee
where we have made use of (\ref{Scat-Amp}), (\ref{A-plus}), and (\ref{ansatz}), and introduced:
    \begin{align}
    &y_{\vec m}:=\left\{\begin{array}{ccc}
    x_{\vec e_0}-2\pi & {\rm for} & \vec m=\vec e_0,\\
    x_{\vec m} & {\rm for} & \vec m\neq\vec e_0,\end{array}\right.
    &&\theta_{\vec m}:={\rm arcsin}\left(\sin\theta_0+\frac{\vec m\cdot\vec\alpha}{k}\right).
    \label{y=}
    \end{align}
According to (\ref{exp-f}), a potential of the form (\ref{multi-exp-v}) scatters a left-incident plane wave with wavenumber $k$ and incident angle $\theta_0$ into a superposition of a finite number of plane waves with wavevector:
	\be
	\bk_{\pm\vec m}=k(\pm\cos\theta_{\vec m}\bbe_x+\sin\theta_{\vec m}\bbe_y).
	\ee

Eqs.~(\ref{exp-f}) and (\ref{y=}) reduce the exact solution of the scattering problem for the potentials (\ref{multi-exp-v}) to inverting the matrix $\bcA$ of coefficients $\cA_{\vec m,\vec n}$ of $\bx_{\vec m}$ in (\ref{sys}). Clearly, by arranging $x_{\vec m}$ and $b_{\vec m}$ into column vectors $\bx_{\vec m}$ and $\bb_{\vec m}$, we can write (\ref{sys}) as $\bcA\bx_{\vec m}=\bb_{\vec m}$. This has a unique solution, namely $\bx_{\vec m}=\bcA^{-1}\bb_{\vec m}$, if and only if $\bcA$ is invertible. The real values of $k$ and $\theta_0$ for which $\bcA$ is a singular matrix correspond to the spectral singularities of the potential (\ref{multi-exp-v}).

Next, we examine the application of our approach for solving the scattering problem for the potential:
	\be
	v(x,y)=\delta(x)\left(\fz_0+\fz_{-}e^{-i\alpha y}+\fz_+ e^{i\alpha y}\right),
	\label{eg1}
	\ee
where $\alpha$ is a positive real parameter, and we use $\fz_\pm$ for what we previously denoted by $\fz_{\pm1}$. In this case, $N=1$ and
	\be
	\cM=\big\{m\in\Z~\big|~ |m \alpha+p_0|\le k~\big\}.
	\label{cM=}
	\ee
In order to determine this set, we consider different ranges of values of $k$ and $p_0$.
	\begin{itemize}
	\item[] Case~1) $\alpha>2k$: Then for all $p_0\in[-k,k]$, (\ref{cM=}) gives $\cM=\{0\}$,
		\begin{align}
		&\cA_{00}=1+\frac{i\fz_0}{2\omega(p_0)},
		&&b_0=2\pi.
		\label{eq-21}
		\end{align}
Using these in (\ref{sys}) we find $x_0=4\pi\omega(p_0)/[2\omega(p_0)+i\fz_0]$, which in view of (\ref{exp-f}) and (\ref{y=}) implies
	\begin{align}
	& f(\theta)=-\frac{\sqrt{2\pi}\fz_0[\delta(\theta-\theta_0)+
	\delta(\theta+\theta_0-\pi)]}{2k\cos\theta_0+i\fz_0}.
	\label{f=11}
	\end{align}
According to this equation, whenever $\fz_0$ takes a positive imaginary value, i.e.,
$\RE(\fz_0)=0$ and $\IM(\fz_0)>0$, and $|\fz_0|<\alpha$, the potential has a spectral singularity at some wavenumber $k\in[\frac{|\fz_0|}{2},\frac{\alpha}{2})$ and incident angle $\theta_0={\rm arccos}(\fz_0/2ik)$.\footnote{For an optical system modeled using the optical potential~(\ref{eg1}), this means that the system emits laser light along the direction given by this value of $\theta_0$, i.e., it acts as a directional laser.} Another interesting consequence of (\ref{f=11}) is that the potentials of the form (\ref{eg1}) with $\fz_0=0$ are omnidirectionally invisible for any left-incident plane wave with wavenumber $k<\alpha/2$.
	\item[] Case~2) $k<\alpha\leq 2k$: Because elements $m$ of $\cM$ satisfy (\ref{condi}), we have $|m|\leq 2k/\alpha$. This together with the fact that $2k/\alpha<2$ imply $\cM\subset\{-1,0,1\}$. Further inspection of the condition $|m \alpha+p_0|\le k$, reveals the following.
	\begin{itemize}
	\item[2.a)] For $|p_0|< \alpha-k$, which corresponds to
	$|\theta_0|<{\rm arcsin}(\alpha/k-1)$, we have $\cM=\{0\}$, and Eqs.~(\ref{eq-21}) and (\ref{f=11}) hold. In particular, if $\fz_0=0$, the potential (\ref{eg1}) is invisible for any left-incident plane wave whose wavenumber $k$ and incident angle $\theta_0$ satisfy $k<\alpha\leq 2k$ and $|\theta_0|<{\rm arcsin}(\alpha/k-1)$.
	
	\item[2.b)] For $p_0\geq \alpha-k$, which corresponds to
	$\theta_0\geq{\rm arcsin}(\alpha/k-1)$, we have $\cM=\{-1,0\}$.
	In this case, (\ref{L-b}) gives
	\begin{align}
	&\bcA=\left[\begin{array}{cc}
	1+\frac{i\fz_0}{2\omega_{-}} & \frac{i\fz_{-}}{2\omega_-}\\[6pt]
	\frac{i\fz_{+}}{2\omega_{0}} & 1+\frac{i\fz_0}{2\omega_0}\end{array}
	\right],
	&&\bb=\left[\begin{array}{c}
	0\\[6pt]
	2\pi\end{array}
	\right],
	\end{align}
where we have employed the shorthand notation:
$\omega_0:=\omega(p_0)$, and $\omega_\pm:=\omega(p_0\pm\alpha)$. It is easy to show that
	\bea
	\det\bcA&=&\frac{\fz_{-}\fz_+-\fz_0^2+2i(\omega_-+\omega_0)\fz_0+4\omega_-\omega_0}{
	4\omega_-\omega_0}\nn\\
	&=&\frac{\fz_{-}\fz_+-\fz_0^2+2ik(\cos\theta_-+\cos\theta_0)\fz_0+4k^2
	\cos\theta_-\cos\theta_0}{
	4k^2\cos\theta_-\cos\theta_0},
	\label{det-A}
	\eea
where $\theta_\pm:={\rm arcsin}(\sin\theta_0\pm\alpha)$.
The zeros of the right-hand side of (\ref{det-A}) gives the spectral singularities of the potential that are located in the range of values of $k$ and $\theta_0$ given by
	\begin{align}
	&k<\alpha\leq 2k, &&{\rm arcsin}(\alpha/k-1)\leq \theta_0\leq\frac{\pi}{2}.
	\label{eq-41}
	\end{align}
For $\det\bcA\neq 0$ we can solve (\ref{sys}) and determine $f(\theta)$ using (\ref{exp-f}). This gives
	\bea
	f(\theta)&=&\frac{\sqrt{2\pi}\,i}{\det\bcA}
	 \left\{\left(\det\bcA-\frac{i\fz_0}{2k\cos\theta_-}-1\right)
	 [\delta(\theta-\theta_0)+\delta(\theta+\theta_0-\pi)]\right.\nn\\
	&&\left.
	\hspace{1.5cm}+\left(\frac{i\fz_-}{2k\cos\theta_-}\right)
	[\delta(\theta-\theta_-)+\delta(\theta+\theta_--\pi)]
	\right\}.
	\label{f=31}
	\eea
According to this relation, the potential (\ref{eg1}) is invisible for incident waves with wavenumber $k$ of incident angle $\theta_0$ satisfying (\ref{eq-41}), if $\fz_0=\fz_-=0$.

	\item[2.c)] For $p_0\leq k-\alpha$, which corresponds to
	$\theta_0\leq-{\rm arcsin}(\alpha/k-1)$, we have $\cM=\{0,1\}$. In this case, a similar analysis leads to (\ref{det-A}) and (\ref{f=31}) with $\theta_-$ changed to $\theta_+$. In particular, spectral singularities of the potential~(\ref{eg1}) at $k$ and $\theta_0$ satisfying $k<\alpha\leq 2k$ and $\theta_0\leq-{\rm arcsin}(\alpha/k-1)$ are given by the zeros of the right-hand side of (\ref{det-A}) with $\theta_-$ changed to $\theta_+$.
		\end{itemize}
	\end{itemize}
	
The above analysis can be extended to obtain a systematic procedure for solving the scattering problem of the potential (\ref{eg1}) for arbitrary $k$ and $\theta_0$.
Let $j$ be the integer part of $2k/\alpha$, i.e., the largest integer that is not larger than $2k/\alpha$. If $j=0$, we have $2k<\alpha$ which we have treated as Case 1 above. For $j\geq 1$, we have
	\be
	\frac{2k}{j+1}<\alpha\leq \frac{2k}{j}.
	\label{condi-3}
	\ee
Setting $j=1$ in this relation, we find Case 2 above.
A close examination of the definition of $\cM$, i.e., the condition $|m\alpha+p_0|\leq k$ on its elements $m$, reveals the fact that whenever (\ref{condi-3}) holds for $j\geq 1$, $\cM$ has either $j$ or $j+1$ elements. The identity of these elements however depends on the value of $p_0$ (alternatively $\theta_0$.) Specifically, for each $q\in\{0,1,2,\cdots,j\}$,
    \be
    \cM=\left\{
    \begin{array}{ccc}
    \{-q,-q+1,-q+2,\cdots,-q+j\} &{\rm for}&-k+q\alpha\leq p_0\leq k-(j-q)\alpha,\\[6pt]
	\{-q,-q+1,-q+2,\cdots,-q+j-1\} &{\rm for}&k-(j-q)\alpha<p_0<-k+(q+1)\alpha.\end{array}\right.
	\ee
Notice that the cases considered in this equation exhausts all possible values of $p_0$, i.e., the interval $[-k,k]$.
	
Having determined $\cM$, we have the range of values of the labels $\ell$ and $m$ of $\cA_{\ell m}$ and $b_\ell$ that specify the system of equations (\ref{sys}). This allows us to determine their explicit form using (\ref{L-b}) and calculate the determinant and inverse of $\bcA$. These respectively give the location of the spectral singularities and the solution of the system, i.e., $x_m$. Substituting the latter in (\ref{y=}) and employing the result in (\ref{exp-f}), we finally find the explicit form of the scattering amplitude.

\subsection{General $y$-periodic potentials of the form $\delta(x)g(y)$}
\label{Sec4-3}

Suppose that $g(y)$ is a periodic function with Fourier series expansion:
	\be
	g(y)=\sum_{n=-\infty}^\infty \fc_n e^{in\alpha y},
	\label{eq-061}
	\ee
where $\fc_n$ and $\alpha$ are respectively complex and positive real parameters. Then (\ref{v=3}) takes the form
	\be
    	v(x,y)=\delta(x)\sum_{n=-\infty}^{\infty}\fz_n e^{in\alpha y},
   	\label{periodic-2}
   	 \ee
where $\fz_n:=\zeta\fc_n$. Substituting (\ref{eq-061}) in (\ref{F=}), we have
	\be
	\cG A_+(p)=-\frac{i}{2\omega(p)}\sum_{n=-\infty}^\infty \fc_n A_+(p-n\alpha).
	\ee
Because $A_+(p)=0$ for $|p|>k$, this equation reduces to (\ref{F=3}), if we identify $N$ with the integer part of $2k/\alpha$, and set $\alpha_n:=n\alpha$. This suggests that we can use the approach of Sec.~\ref{Sec4-2} to solve the scattering problem for the potentials (\ref{periodic-2}). The only difference is that now $N$ is determined by the value of $k$.

The above argument proves the following theorem.
	\begin{itemize}
	\item[]{\em Theorem~1:} Consider a $y$-periodic potential $v(x,y)$ of the form
	$v(x,y):=\delta(x)\sum_{n=-\infty}^\infty\fz_n e^{in\alpha y}$, where $\fz_n$ are complex
	coefficients and $\alpha$ is a positive real parameter. Let $N$ be a non-negative integer, and
	$v_N(x,y):=\delta(x)\sum_{n=-N}^N\fz_n e^{in\alpha y}$. Then the scattering amplitudes of
	$v(x,y)$ and $v_N(x,y)$ coincide for wavenumbers $k<\alpha (N+1)/2$. In particular, if
	$\fz_0=0$, $v(x,y)$ is omnidirectionally invisible for $k<\alpha/2$.
	\end{itemize}

\subsection{Infinite periodic linear array of $\delta$-function potentials  in two dimensions}

Consider the potential
	\be
	v(x,y)=\fz\,\delta(x)\sum_{n=-\infty}^\infty \delta(y-n a),
	\label{pinny2}
	\ee
where $\fz$ is a complex coupling constant and $a$ is a length scale. This potential describes an infinite linear array of equally spaced identical $\delta$-function point interactions. It is a one-dimensional Kronig-Penney potential, also known as a Dirac Comb, embedded in two dimensions. 

Let $\alpha:=2\pi/a$. Then we can use the identity:
	\[\sum_{n=-\infty}^\infty e^{in\alpha y}=a\sum_{n=-\infty}^\infty
	\delta\left( y-an\right),\]
to express (\ref{pinny2}) as
	\be
	v(x,y)=a^{-1}\fz\,\delta(x)\sum_{n=-\infty}^\infty e^{in\alpha y}.
	\label{pinny}
	\ee
Because we can identify (\ref{pinny2}) with (\ref{pinny}), in view of the results of Sec.~\ref{Sec4-3}, we have a complete solution for its scattering problem. In particular, Theorem~1 leads to the following observations.
	\begin{enumerate}
	\item For $k<\alpha/2$, the scattering amplitude for (\ref{pinny2}) is identical to that of $v(x,y)=a^{-1}\fz\,\delta(x)$, i.e., it is given by (\ref{f=11}) with $\fz_0$ replaced with $\fz/a$.
	\item For $\alpha/2\leq k<\alpha$, the scattering amplitude for (\ref{pinny2}) is identical to that of
	\be
	v(x,y)=a^{-1}\fz\,\delta(x)\left(1+e^{i\alpha y}+e^{-i\alpha y}\right)=
	a^{-1}\fz\,\delta(x)[1+2\cos(\alpha y)].\nn
	\ee
Its explicit form are given in our discussion of Case 2 in Sec.~\ref{Sec4-2} with $\fz_0=\fz_\pm=
\fz/a$.
	\end{enumerate}

In Ref.~\cite{linton-martin} the authors consider the scattering problem for an infinite periodic linear array of isotropic point scatterers in two dimensions. They model the point scatterers as closed regions in plane whose size is much smaller than the wavelength of the incident wave. They define the scattering problem by imposing the Dirichlet boundary conditions at the boundary of the scatterers in the limit that the ratio of their size to the wavelength tends to zero. They also demand outgoing boundary conditions for the scattered wave away from the scatterers. Although our approach does not rely on any choice of boundary conditions, there are striking similarities between our results and those of \cite{linton-martin}. For example in both treatments the scattered wave turns out to consist of finitely many plane waves, and the number of these waves as computed in \cite{linton-martin} coincides with the value given by Theorem~1 above. The explicit form of these plane waves, however, differs from those of our approach. To see the reason, we note that the analysis of \cite{linton-martin} makes use of Theorem~1 of Ref.~\cite{KV} which applies to Case I in the classification scheme provided in \cite{KV}. The appearance of $J_0[k(a_m-a_n)]$ in Eq.~(\ref{eq-073}), which has a power series expansion in powers of $k^2$, shows that the scattering problem for the $\delta$-function array (\ref{pinny2}) belongs to Case II of \cite{KV}. This explains the difference between our results and those of \cite{linton-martin}.

\section{Concluding Remarks}

Scattering due to a $\delta$-function potential in one dimension is a standard textbook example of an exactly solvable scattering problem \cite{practical-qm}. Multiple-$\delta$-function potentials in one dimension are also known to lead to exactly solvable scattering problems. In particular, the $n$-th order perturbation theory gives the exact expression for the reflection and transmission amplitudes of a multiple-$\delta$-function potential consisting of $n$ point interactions \cite{pra-2012}. There has also been attempts to explore the scattering properties of semi-infinite periodic arrays of $\delta$-function potentials in one dimension \cite{martin}. The study of higher dimensional generalizations of these potentials, however, is a much more difficult problem \cite{linton-martin,landau-qm,exner}. In particular, the divergences appearing in the Green's function defined by these potentials have led to intricate renormalization schemes for dealing with them \cite{henderson-1997,henderson-1998,Camblong-2001a,camblong-2002}.

Recently we have developed a multi-dimensional transfer matrix formulation of scattering theory that turns out to produce an exact and finite expression for the scattering amplitude of the $\delta$-function potential in two and three dimensions \cite{pra-2016}. In the present article, we have considered a class of scattering problems in two dimensions that share this feature of the $\delta$-function potential in two dimensions. This class includes $\delta$-function potentials that are supported on a subset of a line in two dimensions. Our approach to scattering theory reduces the scattering problem for these potentials to an integral equation that we can solve analytically for a number of interesting cases. Among these are arbitrary finite linear arrays of two-dimensional $\delta$-function potentials, $\delta$-function potentials supported on a line with periodic spatial dependence along the line, and in particular a Dirac comb potential with centers of interaction lying on a line. For all these potentials we offer an analytic scheme for the computation of the exact scattering amplitude. A byproduct of our results is an interesting theorem on potentials that have the same scattering properties in a specific spectral band.

The results we have reported here admit natural extensions to three dimensions. We expect these to be of particular interest in the study of acoustic point scatterers \cite{review,challa}. \\[6pt]

\noindent{\bf Data accessibility statement.} This work does not have any experimental data. \\[6pt]

\noindent{\bf Competing interests statement.} We have no competing interests. \\[6pt]

\noindent{\bf Authors contributions.} Both authors have equally contributed to the formulation of the basic ideas,  conceptual developments, and technical results reported in this paper. AM suggested working on a problem that was later developed by both the authors into the subject of this paper. FL did most of the preliminary calculations. Each author has actively contributed to the improvement of the results initially obtained by the other. AM wrote the paper. Both authors gave final approval for publication. \\[6pt]


\noindent{\bf Funding.} Turkish Academy of Sciences (T\"UBA) has provided the financial support for F. L.'s visit to Ko\c{c} University during which a major part of the research reported here was carried out. This work has been supported by the Scientific and Technological Research Council of Turkey (T\"UB{$\dot{\rm I}$}TAK) in the framework of the Project No.~117F108 and by T\"UBA. \\[6pt]

\noindent{\bf Ethics statement.} This work did not involve any collection of human or animal data.

\ed